\documentclass{article}
\usepackage[T1]{fontenc}
\usepackage{lmodern}
\usepackage{microtype}
\usepackage[margin=1in]{geometry}

\usepackage{graphicx}
\usepackage{subcaption}
\usepackage{amssymb}
\usepackage{nccmath}
\usepackage{bbm}
\usepackage{appendix}
\usepackage{float}
\usepackage{mathtools}

\usepackage{parskip}
\usepackage{xcolor}

\usepackage{natbib}
\usepackage{apalike}
\usepackage[para]{footmisc}
\usepackage{authblk}
\usepackage{amsbsy}

\usepackage{booktabs}
\newtheorem{proposition}{Proposition}
\newtheorem{corollary}{Corollary}

\def\la{\lambda}

\usepackage{hyperref}
\hypersetup{colorlinks=true, allcolors=blue}

\title{Capital allocation on decentralized lending platforms}
\author[1]{Bastien Baude\thanks{\href{mailto:bastien.baude@centralesupelec.fr}{bastien.baude@centralesupelec.fr}}}
\author[2, 3]{Vincent Danos\thanks{\href{mailto:vincent.danos@ens.fr}{vincent.danos@ens.fr}}}
\author[4]{Hamza El Khalloufi\thanks{\href{mailto:hamza.el-khalloufi@univ-paris1.fr}{hamza.el-khalloufi@univ-paris1.fr}}}

\affil[1]{Université Paris-Saclay, CentraleSupélec, 91192 Gif-sur-Yvette, France}
\affil[2]{CNRS, École Normale Supérieure, 45 rue d'Ulm, 75005 Paris, France}
\affil[3]{School of Informatics, University of Edinburgh, Edinburgh EH8 9AB, UK}
\affil[4]{Université Paris 1 Panthéon-Sorbonne, 12 place du Panthéon, 75005 Paris, France}

\begin{document}

\bibliographystyle{apalike}

\maketitle

\begin{abstract}
\noindent
This paper complements \citet{baude2026leveraged}, which studies borrower-side strategies in decentralized lending markets, by focusing on lender-side capital allocation. We consider a lender who seeks to allocate a fixed budget across multiple markets sharing the same supplied asset. Accounting for the impact of supplied capital on lending rates, we derive closed-form solutions under three interest-rate models: linear, kinked, and adaptive (Morpho's AdaptiveCurveIRM). Backtests are conducted first on USDC and then on WETH Morpho lending markets on Ethereum. We also show that, under the kinked rate model, an allocation that brings a market exactly to the kink is never optimal on the lender side, whereas it can be optimal on the borrower side. This asymmetry may create tension between lenders and borrowers around the kink and thereby exacerbate rate volatility.\\


\end{abstract}

\section{Introduction}\label{section:introduction}

This paper is a companion to \citet{baude2026leveraged}, which develops a mathematical framework to optimize leveraged staking, or ``loopy'' strategies on decentralized lending protocols. While \citet{baude2026leveraged} focuses on the borrower side of lending markets, the present work studies the lender side. More precisely, we consider the problem faced by a liquidity supplier who seeks to allocate a given budget across several lending markets sharing the same supplied asset.

We derive closed-form solutions for the optimal allocation problem across multiple lending markets, accounting for the impact of supplied capital on lending rates. These solutions are obtained under three widely used interest-rate models: linear rates, kinked rates (as used by Aave \citep{whitepaper2020aavev1}), and adaptive rates (Morpho's AdaptiveCurveIRM \citep{adaptive2023morpho}). The framework also incorporates market-level allocation limits, treated as lender-specific inputs. These limits may depend on market liquidity, collateral quality and agent behavior, but ultimately reflect the lender's own view of the market. We therefore leave their specification outside the scope of this paper. We then backtest the proposed allocation strategy on real market data from the Ethereum blockchain. The main analysis focuses on USDC markets, and an additional analysis considers WETH markets.

A key difference from \citet{baude2026leveraged} lies in the structure of the optimization problem. On the borrower side, the objective function is concave, whereas on the lender side it is only piecewise concave under the kinked rate model. The solution is therefore obtained by computing a set of candidate solutions and selecting the one that maximizes the objective. The kink also plays a different role on the two sides of the market. We show that an allocation that brings a market exactly at the kink is never optimal on the lender side. This contrasts with the borrower side, where such an allocation can be optimal. This asymmetry may create a persistent tension between lenders and borrowers around the kink and thereby exacerbate rate volatility.

The remainder of the paper is organized as follows. Section \ref{section:mathematical_framework} introduces the main definitions and notation. Section \ref{section:optimal} formulates the allocation problem and derives closed-form solutions under the linear, kinked and adaptive rate models. Section \ref{section:backtest} presents the backtesting results. Section \ref{section:conclusion} concludes and discusses limitations and future research directions.

\section{Supply positions}\label{section:mathematical_framework}

In this preliminary section we fix a few notations pertaining to positions on a single lending market $i$. In the next one, we deal with supply positions that span several lending markets.

From the point of view of a lender, a lending market $i$ can be characterized by:
\begin{itemize}
    \item a liquidity state, with $\bar{S}_{i}$ and $\bar{B}_{i}$ denoting the total value supplied to and borrowed from the market (with $\bar{S}_{i} \ge \bar{B}_{i} > 0$);
    \item an interest rate model mapping the market's utilization rate $\bar{u}_{i} =\bar{B}_{i}/\bar{S}_{i} \leq 1$ to the instantaneous borrow interest rate $b_{i}(\bar{u}_{i})$. The corresponding supply rate is given by $s_{i} = \bar{u}_{i} b_{i}(\bar{u}_{i})$ since the interest paid by borrowers is distributed across all lenders in proportion to their share of supplied liquidity.
\end{itemize}

Suppose we want to take a position $x_{i}$ on market $i$, that is, we deposit an amount $x_{i} \geq 0$ in the lending pool. After our deposit, the total supplied liquidity becomes $\bar{S}_{i} + x_{i}$, while the borrowed amount remains $\bar{B}_{i}$. The interest generated by our position is then given by:
\begin{equation}\label{eq:single_position}
x_{i} s_{i} \Big ( \frac{\bar{B}_{i}}{\bar{S}_{i} + x_{i}} \Big ) = \frac{x_{i} \bar{B}_{i}}{\bar{S}_{i} + x_{i}} b_{i} \Big ( \frac{\bar{B}_{i}}{\bar{S}_{i} + x_{i}} \Big )
\end{equation}
Increasing $x_{i}$ has two opposing effects on the resulting cash flow. On the one hand, it increases our share of the pool, and hence the interest. On the other hand, it decreases the utilization rate, thereby reducing the supply rate earned per unit of supplied capital.

\begin{figure}
    \centering
    \includegraphics[scale = 0.6]{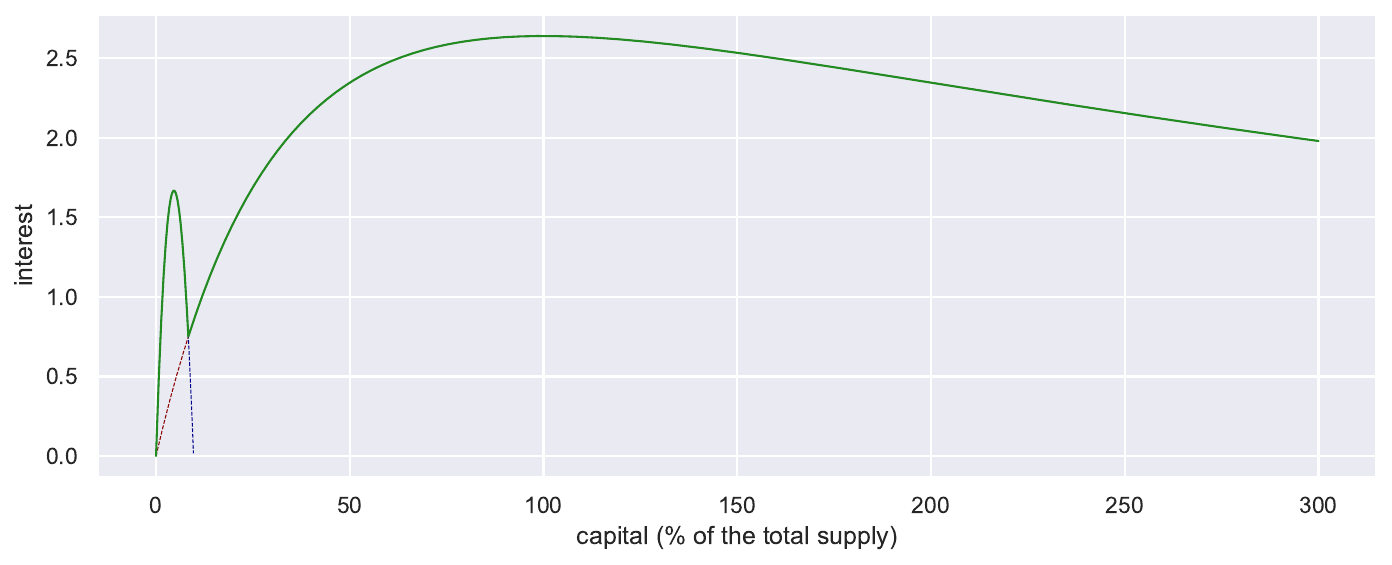}
    \caption{Interest as a function of supplied capital under the kinked model (illustrative example). The dotted lines correspond to the interest generated by the corresponding linear models.}
    \label{fig:non_concave_illustration}
\end{figure}
In contrast to the borrow-side cash flow, which is concave \citep{baude2026leveraged}, the lender-side cash flow \eqref{eq:single_position} is only piecewise concave in $x_{i}$ under the kinked rate model, as illustrated in Figure \ref{fig:non_concave_illustration}.

\section{Optimal capital allocation}\label{section:optimal}

Let $\xi$ denote our total budget in a given asset (e.g., USDC). Our objective is to allocate this budget across $n$ distinct lending markets and an external source of yield so as to maximize total interest. We assume that all lending markets share the same loan asset, namely the budget asset, so that allocations across markets and to the external source of yield are directly comparable and capital can be reallocated between them frictionlessly, apart from gas fees. Our decision variables are the allocations $x_{i}$ to each market $i = 1, \ldots, n$, along with $x_{0}$, our external allocation earning the fixed interest rate $r > 0$. The optimization problem is as follows:
\begin{equation}\label{eq:baseline_problem}
\begin{aligned}
x^{\ast} = \underset{x}{\mathrm{argmax}} \quad & x_{0} r + \sum^{n}_{i=1} x_{i} s_{i} \Big ( \frac{\bar{B}_{i}}{\bar{S}_{i} + x_{i}} \Big ) \\
\textrm{s.t.} \quad & \sum^{n}_{i=0} x_{i} = \xi \\
                    & x_{0} \geq 0 \\
                    & x^{\text{min}}_{i} \leq x_{i} \leq x^{\text{max}}_{i}, \quad i = 1, \ldots, n
\end{aligned}
\end{equation}
Market-level lower and upper allocation limits are included in Problem \eqref{eq:baseline_problem} through $x^{\text{min}}_{i} \leq x_{i} \leq x^{\text{max}}_{i}$ for $i = 1, \ldots, n$. These constraints enable us to incorporate internal risk limits in the allocation decision. However, selecting such limits is not straightforward: it may depend on market liquidity, collateral quality, agent behavior and our own view of the market. We therefore leave this specification outside the scope of this paper. Analogous lower and upper bounds on the external allocation $x_{0}$ are not included in Problem \eqref{eq:baseline_problem} for simplicity; their integration is discussed in Section \ref{section:constraints}.

In what follows, we derive closed-form solutions to Problem \eqref{eq:baseline_problem} under three widely used interest rate models: linear rates, kinked rates (as used by Aave \citep{whitepaper2020aavev1}) and adaptive rates (Morpho's AdaptiveCurveIRM \citep{adaptive2023morpho}).

\subsection{Linear rate}\label{linear_section}

\subsubsection{General solution}

Using the notations introduced in \citet{whitepaper2020aavev1}, the linear model reads:
\begin{equation}\label{eq:linear_rate_model}
b (u) = r_{base} + \frac{u}{u^{*}} r_{slope1}
\end{equation}
where $u^{*} \in (0,1)$ is called the target utilization, $r_{base} \geq 0$ and $r_{slope1} > 0$. At target utilization, the borrow rate is $r_{base} + r_{slope1}$.

We now state its closed-form solution to Problem \eqref{eq:baseline_problem}.
\medskip
\begin{proposition}[Linear rate]\label{pr:linear_rate_solution} Under the linear rate model and assuming $r_{base} < r_{slope1} \frac{\bar{u}_{i}}{u^{*}}$, the optimal allocation of the market-level problem, for a given Lagrange multiplier $\lambda > 0$, is:
\begin{equation}\label{eq:linear_rate_solution}
x^{\ast}_{i}(\lambda) = \min \big ( \max \big ( \kappa(p^{1}_{i}, q^{1}_{i}) - \bar{S}_{i}, x^{\mathrm{min}}_{i} \big ), x^{\mathrm{max}}_{i} \big )
\end{equation}
where,
\begin{equation}\label{eq:roots}
\kappa(p_{i}, q_{i}) = \sqrt[3]{\frac{-q_{i} + \sqrt{\frac{-\Delta(p_{i}, q_{i})}{27}}}{2}} + \sqrt[3]{\frac{-q_{i} - \sqrt{\frac{-\Delta(p_{i}, q_{i})}{27}}}{2}}
\end{equation}
and the discriminant $\Delta$ is given by:
\begin{equation}
\Delta(p_{i}, q_{i}) = - \big ( 4 p^{3}_{i} + 27 q^{2}_{i} \big )
\end{equation}
with,
\begin{equation}\label{eq:q_p_1}
p^{1}_{i} = \frac{\bar{B}_{i} \bar{S}_{i}}{\lambda} \Big ( r_{slope1} \frac{\bar{u}_{i}}{u^{*}} - r_{base} \Big ), \quad q^{1}_{i} = - 2 r_{slope1} \frac{\bar{B}_{i}^{2} \bar{S}_{i}}{\lambda u^{*}}
\end{equation}
We then distinguish two cases.

\paragraph{Saturated markets \normalfont ($\lambda^{\ast} = r$)} For each market $i$, we determine the optimal allocation $x^{\ast}_{i}(r)$ given by \eqref{eq:linear_rate_solution}. The optimal external allocation $x^{\ast}_{0}$ follows from the budget constraint:
\begin{equation}\label{eq:x_0_saturated}
x^{\ast}_{0} = \xi - \sum^{n}_{i=1} x^{\ast}_{i}(r)
\end{equation}
In this regime, all markets are saturated,\footnote{meaning, had we more budget it would be advantageous to lend more} and the residual budget is allocated to the external source of yield.

\paragraph{Unsaturated markets \normalfont ($\lambda^{\ast} > r$)} In this regime, the total budget is not sufficient to saturate all markets, so the external allocation vanishes ($x^{\ast}_{0} = 0$). Moreover, for each market $i$, the optimal allocation $x^{\ast}_{i}(\lambda)$ is given by \eqref{eq:linear_rate_solution} and the optimal Lagrange multiplier $\la^{\ast}$ is the solution to the budget constraint:
\begin{equation}\label{eq:budget_constraint_unsaturated}
\sum^{n}_{i=1} x^{\ast}_{i}(\la^{\ast}) = \xi
\end{equation}
\end{proposition}
The proof is provided in Appendix \ref{appendix:proof_linear_rate_solution}.

\subsubsection{Algorithm}\label{algorithm}

We first compute the optimal solution \eqref{eq:linear_rate_solution} for each market wih $\lambda = r$, and denote the solution by $x^{\ast, (0)}$. If $\sum^{n}_{i=1} x^{\ast, (0)}_{i} \leq \xi$, the solution is valid and we determine the external allocation using $x^{\ast, (0)}_{0} = \xi - \sum^{n}_{i=1} x^{\ast, (0)}_{i}$.

Otherwise, we solve \eqref{eq:linear_rate_solution} for each market given $\lambda > r$, and denote the resulting solution by $x^{\ast, (1)}(\lambda)$ for which $x^{\ast, (1)}_{0} = 0$. The optimal multiplier $\lambda^{\ast}$ is then solve numerically using Brent’s algorithm to satisfy:
\begin{equation}\label{eq:f}
\sum^{n}_{i=1} x^{\ast,(1)}_{i}(\lambda^{\ast}) = \xi
\end{equation}
thereby ensuring that the allocation satisfies the budget constraint.

\subsubsection{Interpreting $\la^{\ast}$}

One can think of $\la^{\ast}$, the optimal Lagrange multiplier, as a way to make the external source of yield artificially and gradually more profitable, up to the point where all lending markets become saturated under the budget constraint.

\subsection{Kinked rate}\label{kinked_section}

\subsubsection{General solution}

The kinked model reads:
\begin{equation}\label{eq:kinked_rate_model}
b(u) = \left\{
    \begin{array}{ll}
        \displaystyle r_{base} + \frac{u}{u^{*}} r_{slope1} & \mbox{if } u <  u^{*} \\
        \displaystyle r_{base} + r_{slope1} + \frac{u - u^{*}}{1 - u^{*}} r_{slope2} & \mbox{if } u \geq u^{*}
    \end{array}
\right.
\end{equation}
where $u^{*}\in(0,1)$, $r_{base} \geq 0$, $r_{slope1} > 0$ and $r_{slope2} > 0$. Again, we use the notation of \citet{whitepaper2020aavev1}. The parameter $r_{slope1}$ is normalized such that the rate at target utilization $u^{*}$ equals $r_{base} + r_{slope1}$, while $r_{slope2}$ is normalized so that the rate at full utilization equals $r_{base} + r_{slope1} + r_{slope2}$.

We now state the closed-form solution to Problem \eqref{eq:baseline_problem} under the kinked rate model.
\medskip
\begin{proposition}[Kinked rate]\label{pr:kinked_rate_solution} Under the kinked rate model \eqref{eq:kinked_rate_model} and assuming $r_{slope1} \frac{\bar{u}_{i}}{u^{*}} > r_{base}$ and $r_{slope2} > (r_{base} + r_{slope1}) \frac{1 - u^{*}}{2 u^{*}}$, the market-level candidate allocation, for a given Lagrange multiplier $\lambda > 0$, reads as follows. If $\bar{u}_{i} \geq u^{*}$, i.e., when the current utilization is above the target rate, the market-level candidate allocation is:
\begin{equation}\label{eq:kinked_rate_solution_below}
x^{*}_{i}(\lambda) = \left\{
    \begin{array}{ll}
        \min \big ( \max \big ( \kappa(p^{1}_{i}, q^{1}_{i}) - \bar{S}_{i}, x^{\mathrm{min}}_{i} \big ), x^{\mathrm{max}}_{i} \big ) & \mbox{if } x^{*}_{i} \geq x^{(\mathrm{kink})}_{i} \\
        \min \big ( \max \big ( \kappa(p^{2}_{i}, q^{2}_{i}) - \bar{S}_{i}, x^{\mathrm{min}}_{i} \big ), x^{\mathrm{max}}_{i} \big ) & \mbox{if } x^{*}_{i} < x^{(\mathrm{kink})}_{i}
    \end{array}
\right.
\end{equation}
where $x^{(\mathrm{kink})}_{i} = \frac{\bar{B}_{i}}{u^{*}} - \bar{S}_{i}$. Additionally, $p^{1}_{i}, q^{1}_{i}$ are given by \eqref{eq:q_p_1} and:
\begin{equation}
p^{2}_{i} = \frac{\bar{B}_{i} \bar{S}_{i}}{\lambda} \Big ( r_{slope2} \frac{\bar{u}_{i} + u^{*}}{1 - u^{*}} - r_{base} - r_{slope1} \Big ), \quad q^{2}_{i} = - 2 r_{slope2} \frac{\bar{B}_{i}^{2} \bar{S}_{i}}{\lambda (1 - u^{*})}
\end{equation}
Otherwise, if $\bar{u}_{i} < u^{*}$, i.e., when the current utilization is below the target rate, the market-level candidate solution is given by \eqref{eq:linear_rate_solution}.
\end{proposition}

\begin{corollary}[Non-optimality at the kink]\label{co:kink_non_optimal} Assuming $\frac{r_{slope2}}{1-u^{*}} > \frac{r_{slope1}}{u^{*}}$, the allocation $x^{(\mathrm{kink})}_{i}$ is never optimal for the market-level problem.
\end{corollary}
The proof is provided in Appendix \ref{appendix:proof_kinked_rate_solution}. In practice, the condition $\frac{r_{slope2}}{1-u^{*}} > \frac{r_{slope1}}{u^{*}}$, i.e., the slope of the borrow rate above the target utilization is steeper than the slope below, is always satisfied.

\subsubsection{Algorithm}

Because the objective in Problem \eqref{eq:baseline_problem} is only piecewise concave under the kinked rate model, we enumerate the admissible regime configurations and retain the feasible allocation that maximizes the objective. For each market $i$, we define the set of admissible regimes $\mathcal{M}_{i}$ by:
\begin{equation}\label{eq:modes}
\mathcal{M}_{i} = \left\{
    \begin{array}{ll}
        \{1\} & \text{if } \bar{u}_{i} < u^{*} \\
        \{1,2\} & \text{if }  \bar{u}_{i} \ge u^{*} \text{ and } x_{i}^{\mathrm{(kink)}} \le \xi \text{ and } \lambda_{i}^{1} \geq r \\
        \{2\} & \text{otherwise}
    \end{array}
\right.
\end{equation}
where,
\begin{equation}
\lambda^{1}_{i} = u^{*} \Big (\frac{u^{*}}{\bar{u}_{i}} (r_{base} + r_{slope1}) + r_{slope1} \big ( \frac{u^{*}}{\bar{u}_{i}} - 1 \big ) \Big )
\end{equation}
Here, regime $1$ denotes the pre-kink region, and regime $2$ denotes the post-kink region. The reasoning is as follows. When the current utilization is below the target, the market already lies in the pre-kink region, so only regime $1$ has to be considered. When the current utilization is above the target, regime $2$ is always admissible. Regime $1$, however, is considered only if two additional conditions hold. First, the budget must be large enough to reach the kink, that is, $x_{i}^{\mathrm{(kink)}} \le \xi$. Second, reaching the pre-kink region must be profitable relative to the external source of yield, which is captured by the condition $\lambda_{i}^{1} \ge r$. Using Corollary \ref{co:kink_non_optimal}, the kink regime can be disregarded.

A regime scenario is a vector:
\begin{equation}
m = (m_1, \ldots, m_n) \in \mathcal{M}_{1} \times \cdots \times \mathcal{M}_{n}
\end{equation}
Additionally, a regime scenario $m$ is feasible only if:
\begin{equation}
\sum_{i=1}^{n} x_{i}^{\mathrm{(kink)}} \mathbbm{1}_{\{ m_{i} = 1, \ \bar{u}_{i} \ge u^{*} \}} \le \xi
\end{equation}
Otherwise, the budget is not sufficient to reach the pre-kink regime in all markets selected by $m$, and the scenario is discarded.

\subsection{Adaptive rate}\label{adaptive_section}

As mentioned in \citet{baude2026leveraged}, after omitting the dynamic feature of the adaptive rate---which is justified in our case since the optimization problem is static in time---the adaptive model is a reparametrization of the kinked model \eqref{eq:kinked_rate_model}. The corresponding parameters are:
\begin{equation}\label{eq:mapping_adaptive_to_kinked}
r_{base} = \frac{r^{\text{target}}_{t}}{k_{d}}, \quad r_{slope1} = r^{\text{target}}_{t} \big ( 1 - \frac{1}{k_{d}} \big ), \quad r_{slope2} = r^{\text{target}}_{t} \big ( k_{d} - 1 \big )
\end{equation}
Under the adaptive parametrization \eqref{eq:mapping_adaptive_to_kinked}, the assumptions of Proposition \ref{pr:kinked_rate_solution} reduce to: $\bar{u}_{i} > \frac{u^{*}}{k_{d} - 1}$ and $k_{d} > \frac{1 + u^{*}}{2 u^{*}}$. In practice, the values $u^{*} = 0.9$ and $k_{d} = 4$ are hardcoded in the smart contract. The second condition is therefore always satisfied, and the first one holds whenever $\bar{u}_{i} > 30 \%$. In mature markets, violations of this condition are rare and occur only under severe liquidity stress scenarios.

\subsection{Dealing with allocation limits on the external source of yield}\label{section:constraints}

In Problem \eqref{eq:baseline_problem}, we only imposed a non-negativity constraint on the allocation $x_{0}$ to the external source of yield. In some applications, however, it may be necessary to impose lower and upper allocation limits. We therefore consider constraints of the form $x^{\text{min}}_{0} \leq x_{0} \leq x^{\text{max}}_{0}$. The KKT conditions distinguish three cases: if $\lambda > r$, $x^{*}_{0} = x^{\text{min}}_{0}$, if $\lambda < r$, $x^{*}_{0} = x^{\text{max}}_{0}$ else if $\lambda = r$, $x^{*}_{0} \in [x^{\text{min}}_{0}, x^{\text{max}}_{0}]$. The lower bound is therefore naturally incorporated by reserving this amount for the external source and solving the allocation problem over the remaining budget. The upper bound, however, is more involved. Once it is imposed, the optimum may occur in a regime where the Lagrange multiplier satisfies $\lambda < r$. As a result, we can no longer guarantee that $\lambda > 0$, a property used to derive the market-level closed-form solution. The upper bound on the allocation $x_{0}$ therefore requires a separate analysis, which is not treated in this paper.

\section{Numerical results}\label{section:backtest}

\subsection{Data processing}\label{section:data}

\begin{figure}
    \centering
    \begin{subfigure}{\linewidth} 
        \centering
        \includegraphics[scale = 0.6]{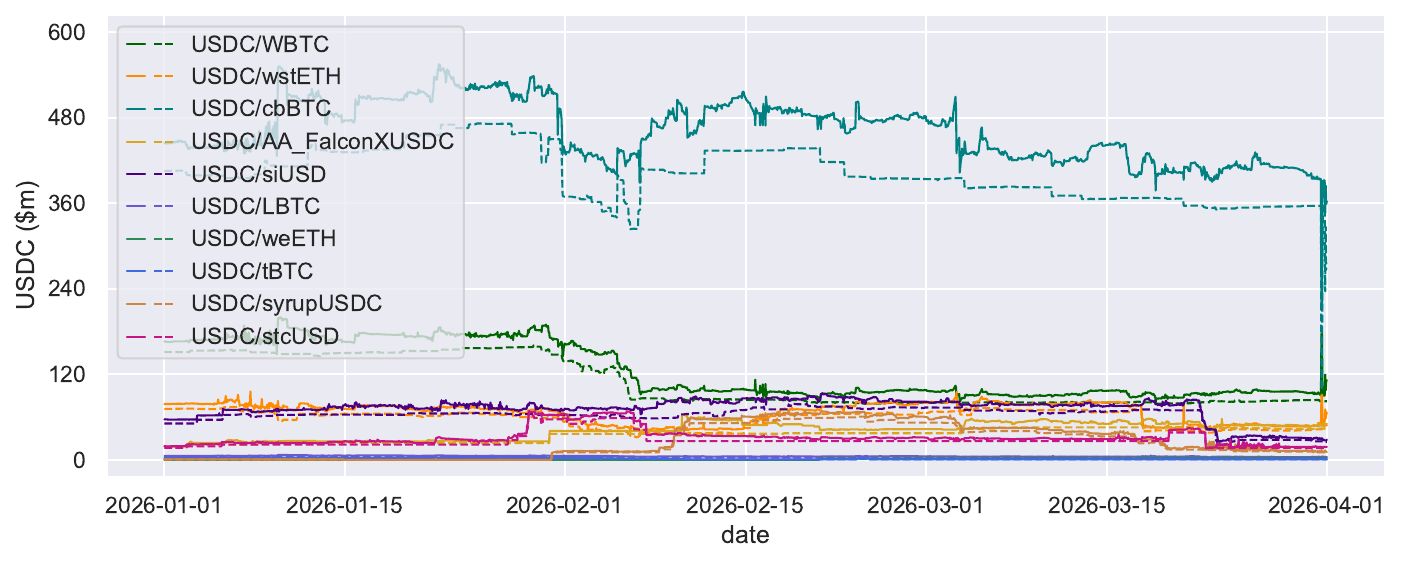}
        \caption{reserves}
        \label{fig:evolution_core_market_reserves}
    \end{subfigure}
    \vfill
    \begin{subfigure}{\linewidth}
        \centering
        \includegraphics[scale = 0.6]{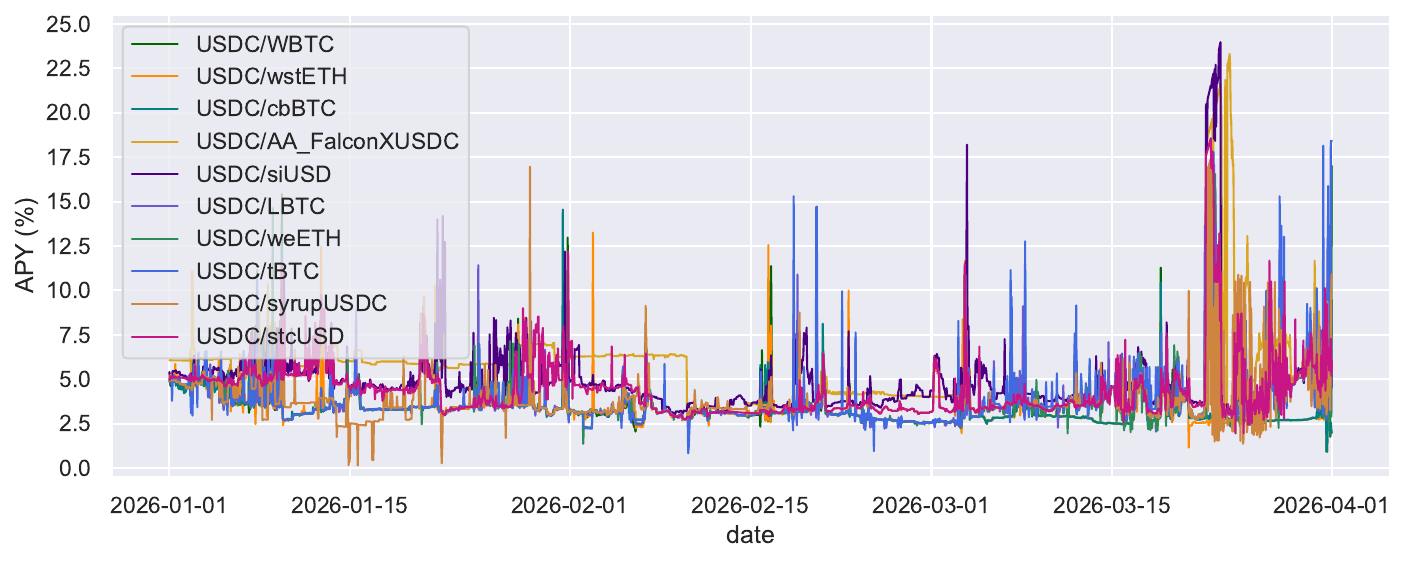}
        \caption{interest rate}
        \label{fig:evolution_core_market_rates}
    \end{subfigure}
    \caption{Evolution of USDC reserves (solid line: supplied funds; dashed line: borrowed fund) and the supply rate of the core-inspired markets on Morpho on the Ethereum blockchain from January 1, 2026 to April 1, 2026.}
    \label{fig:evolution_core_market}  
\end{figure}

\begin{table}
\centering
\begin{tabular}{ccc}
\toprule
market & ID & creation date \\
\midrule
USDC/cbBTC & \texttt{4f598145-0188-44dc-9e18-38a2817020a1} & September 4, 2024 \\
USDC/WBTC & \texttt{7ce50eac-4df6-440c-a374-a446e483669c} & January 17, 2024 \\
USDC/wstETH & \texttt{c9916204-db88-4ac0-999f-0bba63957dc4} & January 3, 2024 \\
\bottomrule
\end{tabular}
\caption{IDs and creation dates of the prime-inspired markets on Morpho on the Ethereum blockchain.}
\label{tab:prime_markets}
\end{table}
\begin{figure}
    \centering
    \begin{subfigure}{\linewidth} 
        \centering
        \includegraphics[scale = 0.6]{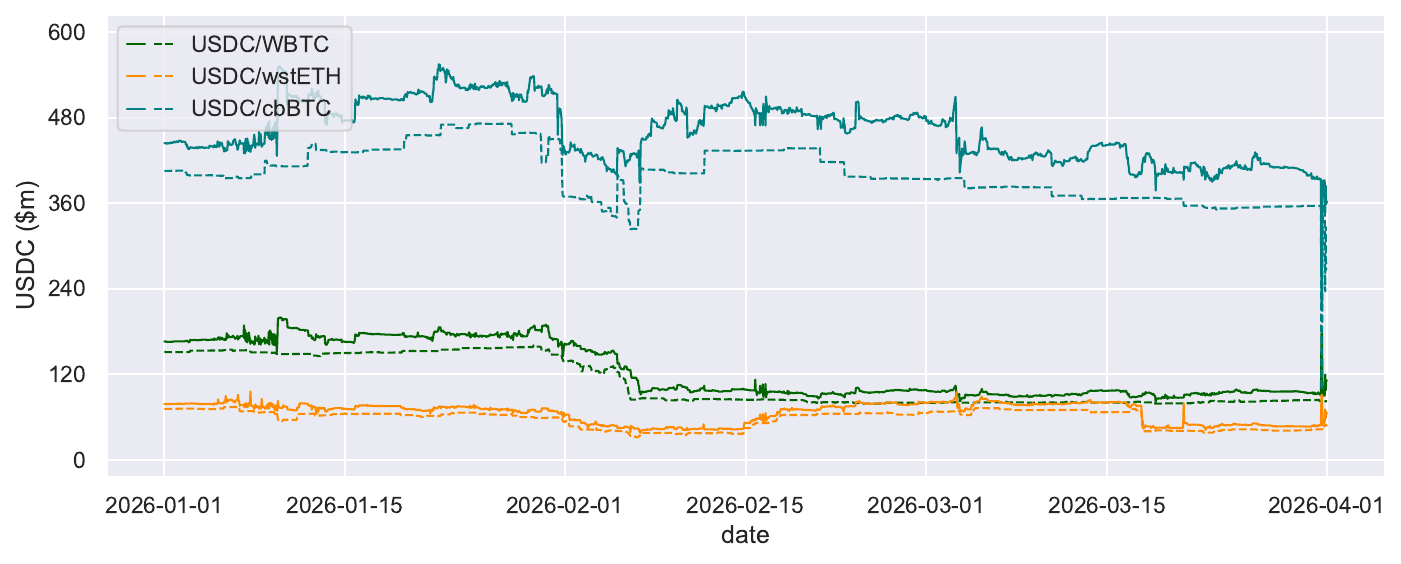}
        \caption{reserves}
        \label{fig:evolution_prime_market_reserves}
    \end{subfigure}
    \vfill
    \begin{subfigure}{\linewidth}
        \centering
        \includegraphics[scale = 0.6]{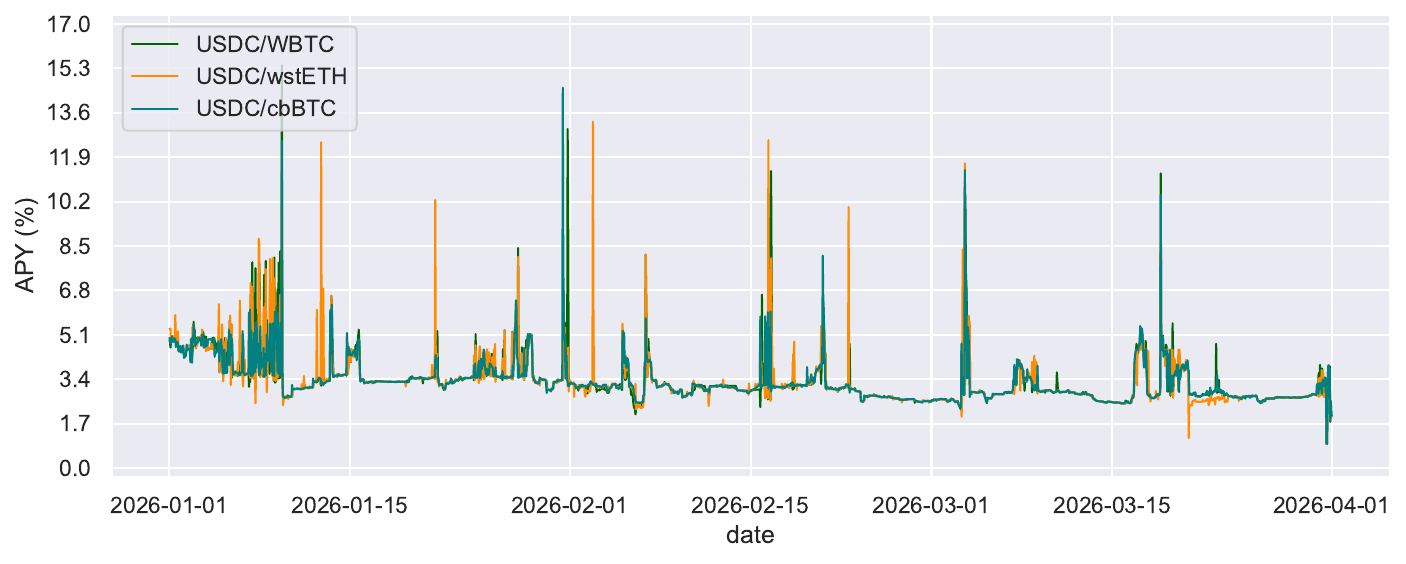}
        \caption{interest rate}
        \label{fig:evolution_prime_market_rates}
    \end{subfigure}
    \caption{Evolution of USDC reserves (solid line: supplied funds; dashed line: borrowed fund) and the supply rate of the prime-inspired markets on Morpho on the Ethereum blockchain from January 1, 2026 to April 1, 2026.}
    \label{fig:evolution_prime_market}
\end{figure}
The strategy is backtested on real market data from the Ethereum blockchain. We first describe the dataset and preprocessing steps before turning to the results. We focus on USDC markets on Morpho over the period from January 1, 2026, to April 1, 2026. The data is retrieved from Morpho's GraphQL service.\footnote{\url{https://api.morpho.org/graphql}} Specifically, we consider two sets of markets:
\begin{itemize}
\item \textbf{prime-inspired markets}: we select the three markets included in the Gauntlet USDC Prime vault;\footnote{vault address: \texttt{0xdd0f28e19C1780eb6396170735D45153D261490d}}
\item \textbf{core-inspired markets}: we select ten markets in which the Gauntlet USDC Core vault\footnote{vault address: \texttt{0x8eB67A509616cd6A7c1B3c8C21D48FF57df3d458}} has already allocated capital. Markets collateralized by Pendle PT tokens are excluded because the collateral has a fixed maturity. Maintaining such positions would require rolling them into markets with later maturities, which is outside the scope of this paper.
\end{itemize}
The characteristics of these markets are summarized in Tables \ref{tab:prime_markets}, \ref{tab:core_markets}. Figures \ref{fig:evolution_prime_market_reserves}, \ref{fig:evolution_core_market_reserves} show the evolution of supplied and borrowed USDC in the selected prime-inspired and core-inspired markets, respectively. Figures \ref{fig:evolution_prime_market_rates}, \ref{fig:evolution_core_market_rates} show the corresponding evolution of supply interest rates over the same period. To reduce noise in the backtest, we use one-day moving averages of the market variables as inputs to the allocation algorithm.
\begin{table}
\centering
\begin{tabular}{ccc}
\toprule
market & ID & creation date \\
\midrule
USDC/cbBTC & \texttt{4f598145-0188-44dc-9e18-38a2817020a1} & September 4, 2024 \\
USDC/WBTC & \texttt{7ce50eac-4df6-440c-a374-a446e483669c} & January 17, 2024 \\
USDC/wstETH & \texttt{c9916204-db88-4ac0-999f-0bba63957dc4} & January 3, 2024 \\
USDC/AA\_FalconXUSDC & \texttt{5d492ce2-295a-47ac-bf20-843a2fa0bb49} & July 3, 2025 \\
USDC/siUSD & \texttt{2802d9a9-29b7-4202-8f40-38a44a395d4b} & July 11, 2025 \\
USDC/stcUSD & \texttt{acaff9b4-7cbc-4251-9027-473a7ef9004d} & August 28, 2025 \\
USDC/LBTC & \texttt{7f9975ae-6f3c-4fa6-8f47-96bc9c88418b} & January 29, 2025 \\
USDC/weETH & \texttt{f9973074-bebd-47c9-9f4f-69d681d888ec} & September 23, 2024 \\
USDC/tBTC & \texttt{7e34b818-199b-4cdf-b789-a89bedd288ac} & May 29, 2024 \\
USDC/syrupUSDC & \texttt{60e7a954-0057-43e9-a190-f7ed73eeab4b} & March 9, 2025 \\
\bottomrule
\end{tabular}
\caption{IDs and creation dates of the core-inspired markets on Morpho on the Ethereum blockchain.}
\label{tab:core_markets}
\end{table}

\subsection{Backtesting}\label{section:backtest_results}

\begin{table}
\centering
\begin{tabular}{ccc}
\toprule
strategy & initial investment ($\$$) & APY ($\%$) \\
\midrule
prime-inspired (low cap) & $100 \text{k}$ & $3.3$ \\
core-inspired (low cap) & $100 \text{k}$ & $5.5$ \\
USDC/cbBTC (low cap) & $100 \text{k}$ & $3.3$ \\
prime-inspired (high cap) & $100 \text{m}$ & $2.4$ \\
core-inspired (high cap) & $100 \text{m}$ & $2.8$ \\
USDC/cbBTC (high cap) & $100 \text{m}$ & $2.2$ \\
\bottomrule
\end{tabular}
\caption{Performance of the strategy compared to USDC/cbBTC on Morpho on the Ethereum blockchain from January 1, 2026 to April 1, 2026.}
\label{tab:ethereum_strategies}
\end{table}
We backtest the strategy using the dataset described above. Two budget configurations are considered: \$100k, corresponding to a low-impact allocation, and \$100m, corresponding to a high-impact allocation. The strategy is rebalanced daily. Table \ref{tab:ethereum_strategies} reports the resulting APYs over the backtesting period. As a benchmark, we include a passive strategy that allocates the entire budget to the most liquid market, namely USDC/cbBTC. The core-inspired strategy outperforms both the prime-inspired one and the USDC/cbBTC benchmark because its market set contains more attractive lending opportunities. As expected, high-capital strategies generate lower APYs than their low-capital counterparts. This is due to the \textit{dilution effect}: larger allocations reduce utilization and therefore lower the supply rates. Additionally, in the low-capital case, the prime-inspired strategy and the USDC/cbBTC benchmark achieve similar APYs because the \textit{dilution effect} is limited and the rates of the prime-inspired markets are close to the USDC/cbBTC rate. In the high-capital case, the \textit{dilution effect} affects both the prime-inspired strategy and the benchmark. However, it is less pronounced for the prime-inspired strategy, whose aggregate liquidity is larger than the benchmark's, resulting in a higher performance.

\begin{figure}
    \centering
    \begin{subfigure}{\linewidth} 
        \centering
        \includegraphics[scale = 0.6]{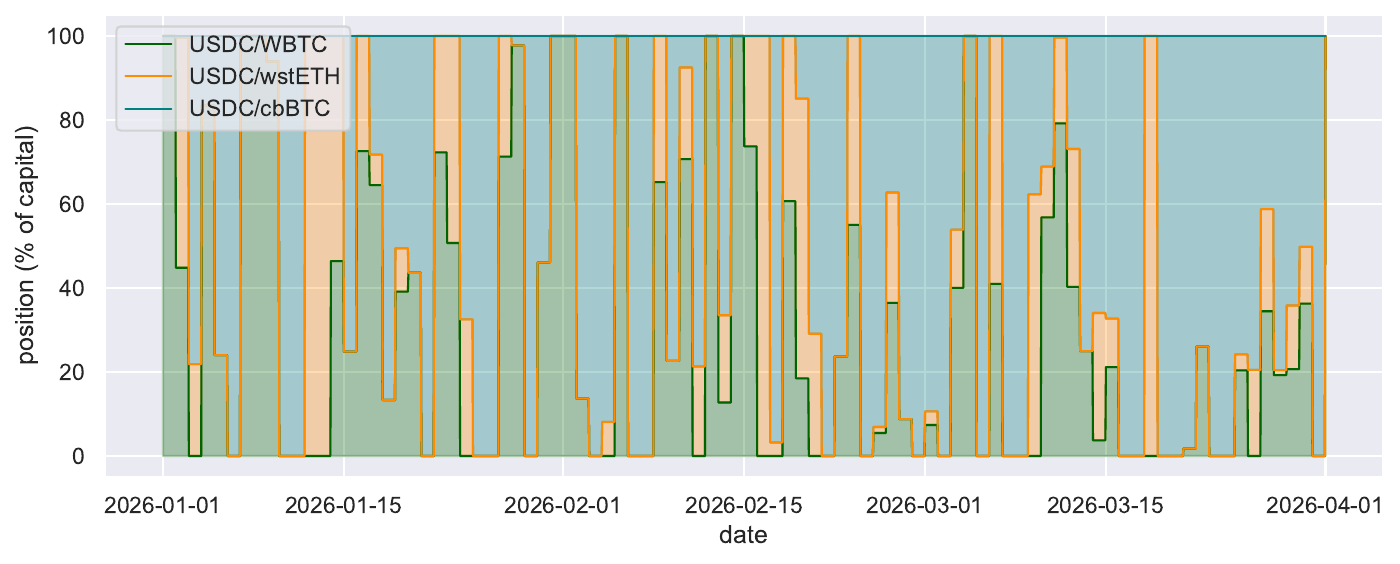}
        \caption{low cap}
    \end{subfigure}
    \vfill
    \begin{subfigure}{\linewidth}
        \centering
        \includegraphics[scale = 0.6]{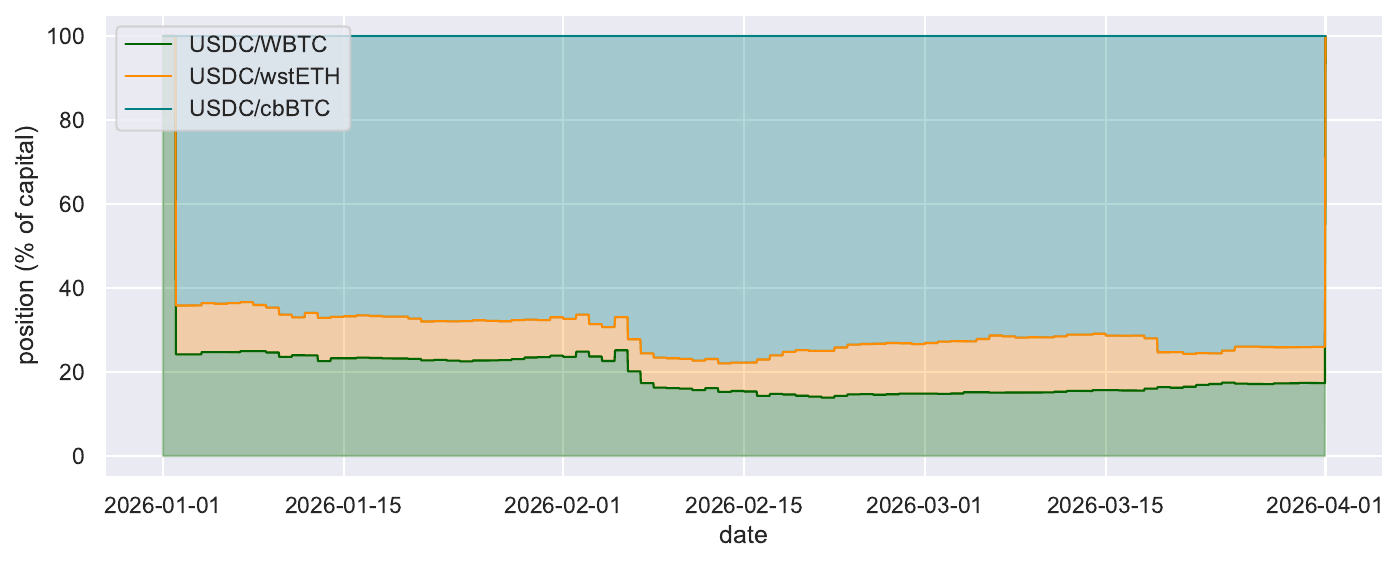}
        \caption{high cap}
    \end{subfigure}
    \caption{Evolution of the positions of the prime-inspired strategy on Morpho on the Ethereum blockchain from January 1, 2025 to April 1, 2025.}
    \label{fig:backtesting_ethereum_prime}    
\end{figure}
\begin{figure}
    \centering
    \begin{subfigure}{\linewidth} 
        \centering
        \includegraphics[scale = 0.6]{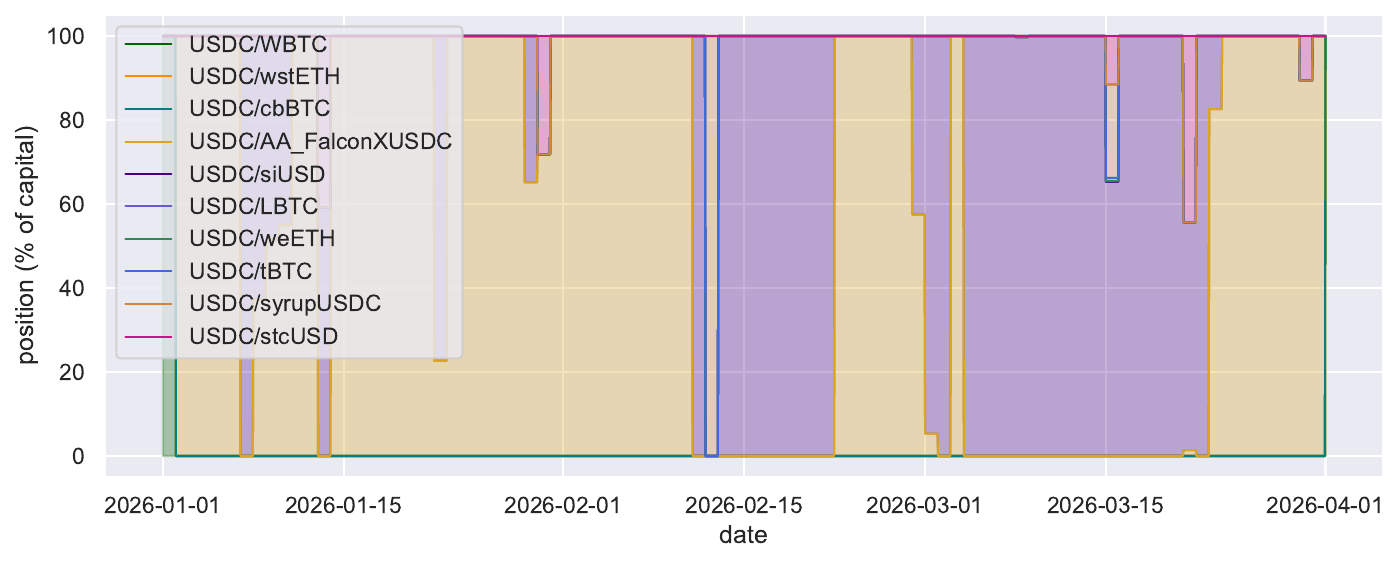}
        \caption{low cap}
    \end{subfigure}
    \vfill
    \begin{subfigure}{\linewidth}
        \centering
        \includegraphics[scale = 0.6]{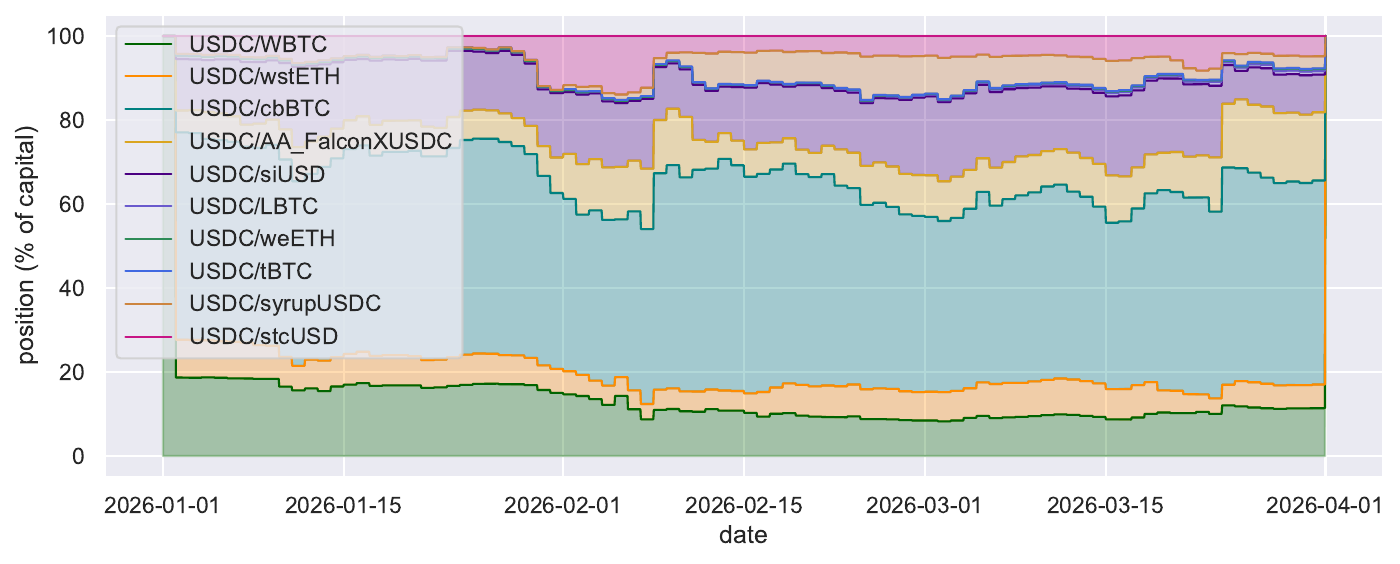}
        \caption{high cap}
    \end{subfigure}
    \caption{Evolution of the positions of the core-inspired strategy on Morpho on the Ethereum blockchain from January 1, 2025 to April 1, 2025.}
    \label{fig:backtesting_ethereum_core}    
\end{figure}

Figures \ref{fig:backtesting_ethereum_prime}, \ref{fig:backtesting_ethereum_core} show the evolution of the optimal allocations over the backtesting period for the prime-inspired and core-inspired strategies, respectively. For the low-capital strategy on the core-inspired markets, the allocation exhibits a \textit{bang-bang} pattern: the optimal position switches discontinuously from one market to another. This behavior is much less pronounced for the prime-inspired markets. The difference stems from the behavior of supply rates across the two market sets. In the prime-inspired set, rates tend to move together and often converge toward a common level, except during short-lived liquidity stress episodes. This convergence suggests that market participants view the collateral risks across these markets as homogeneous. As a result, when rates are sufficiently close, even the low-impact budget should be split across those markets. By contrast, rate spreads across the core-inspired markets are significant and persistent, reflecting the heterogeneity of collateral risks priced by the market. With a low-impact budget, the strategy can exploit these spreads by investing almost exclusively in the most profitable market, without significantly affecting market utilization. This leads to the observed \textit{bang-bang} behavior. For the high-capital case, in both market sets, the allocation displays allocation bands. These bands are closely related to the liquidity available in each market: larger and more liquid markets receive larger allocations, while high-yield opportunities with limited liquidity vanish quickly.

\begin{figure}
    \centering
    \includegraphics[scale = 0.6]{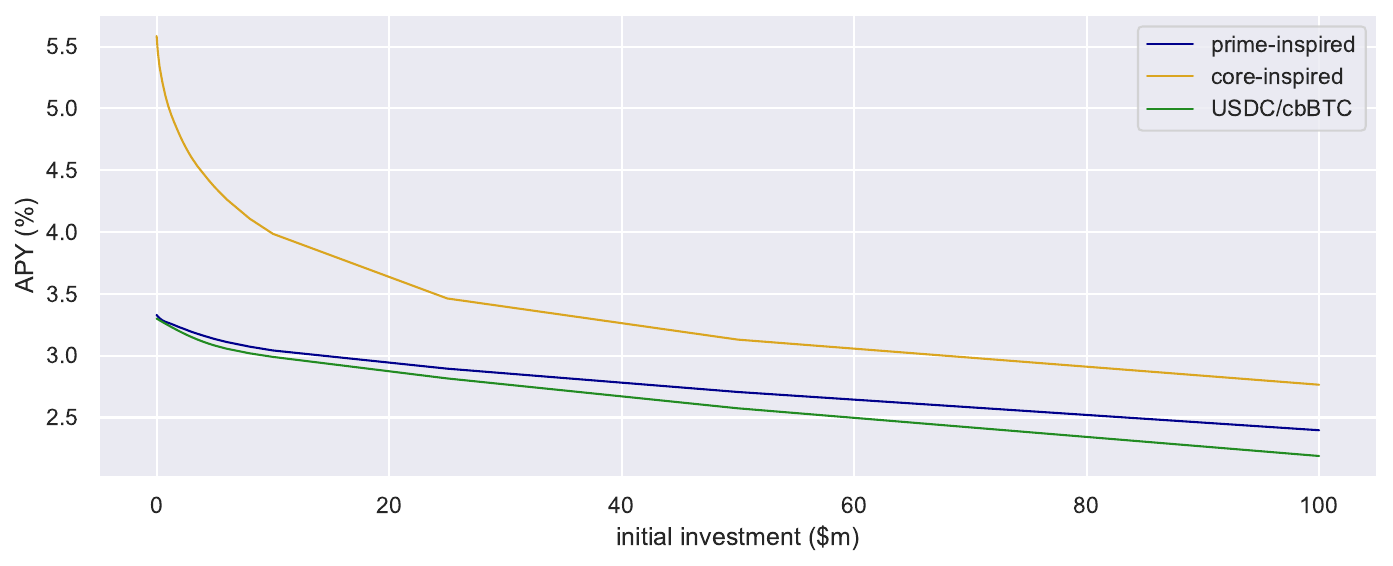}
    \caption{APY of the strategy with respect to initial investment on Morpho on the Ethereum blockchain from January 1, 2026 to April 1, 2026.}
    \label{fig:backtesting_ethereum_apys}
\end{figure}

Figure \ref{fig:backtesting_ethereum_apys} presents the APY as a function of the initial investment. For all strategies, APY decreases with investment size because of the \textit{dilution effect}. For small investments, the core-inspired market set significantly outperforms the other strategies because it allocates capital to more attractive markets with limited liquidity. As the investment size increases, these markets are progressively exhausted, and their rates converge toward the main market rates. As a result, the performance gap between strategies narrows. Beyond a certain investment size, once all market rates have converged to a common rate, the APY behaves as if capital were allocated to a single market with aggregated liquidity.

(See Appendix \ref{appendix:weth} for the backtest on WETH markets.)

\subsection{Local maxima under the kinked rate model}

\begin{table}
\centering
\begin{tabular}{ccccccc}
\toprule
market & $u^{*}$ (\%) & $r_{base}$ (\%) & $r_{slope1}$ (\%) & $r_{slope2}$ (\%) & $\bar{B}$ & $\bar{S}$ \\
\midrule
$1$ & $90$ & $0$ & $5$ & $17.8$ & $923$ & $1000$ \\
$2$ & $90$ & $0$ & $5$ & $16.6$ & $936$ & $1000$ \\
$3$ & $90$ & $0$ & $5$ & $20.0$ & $920$ & $1000$ \\
$4$ & $90$ & $0$ & $5$ & $18.7$ & $924$ & $1000$ \\
$5$ & $90$ & $0$ & $5$ & $20.0$ & $943$ & $1000$ \\
\bottomrule
\end{tabular}
\caption{Interest-rate parameters and liquidity state of the five synthetic markets.}
\label{tab:market_params}
\end{table}
We compare our closed-form allocation with two SLSQP-based numerical benchmarks:
\begin{itemize}
\item \textbf{SLSQP}: the solver is initialized with the full budget allocated to the external source of yield;
\item \textbf{multi-start SLSQP}: the solver is run from several initial allocations, and the best resulting candidate is retained. The initial allocations consist of the full budget allocated to the external source of yield, the full budget allocated to each market in turn, and uniform allocations across markets, either including or excluding the external source of yield.
\end{itemize}

\begin{table}
\centering
\begin{tabular}{ccc}
\toprule
algorithm & APY (\%) & time (s) \\
\midrule
SLSQP & $3.0$ & $0.01$ \\
multi-start SLSQP & $4.3$ & $0.04$ \\
closed-form & $4.6$ & $0.04$ \\
\bottomrule
\end{tabular}
\caption{Performance across optimization algorithms.}
\label{tab:performance}
\end{table}
We start with a set of five synthetic lending markets; their characteristics are reported in Table \ref{tab:market_params}. The APY and computation time obtained with each algorithm are reported in Table \ref{tab:performance}. SLSQP-based methods are trapped in local maxima, while our method outperforms both, with a computation time comparable to multi-start SLSQP. The corresponding allocations are reported in Table \ref{tab:allocations}.
\begin{table}
\centering
\begin{tabular}{ccccccc}
\toprule
algorithm & $x_{0}$ & $x_{1}$ & $x_{2}$ & $x_{3}$ & $x_{4}$ & $x_{5}$ \\
\midrule
SLSQP & $161.7$ & $24.8$ & $32.7$ & $21.8$ & $24.8$ & $34.2$ \\
multi-start SLSQP & $0$ & $56.6$ & $63.8$ & $54.9$ & $57.1$ & $67.6$ \\
closed-form & $0$ & $83.6$ & $23.6$ & $81.9$ & $84.2$ & $26.6$ \\
\bottomrule
\end{tabular}
\caption{Optimal allocations across optimization algorithms.}
\label{tab:allocations}
\end{table}

We also compare the results obtained with the three methods for the backtests performed in Section \ref{section:backtest_results}. The results are nearly identical. This suggests that, in practice, the potential differences between the closed-form algorithm and the numerical ones are largely drowned out by market noise.

\section{Discussion and conclusion}\label{section:conclusion}

We propose a mathematical framework for optimal capital allocation on decentralized lending markets. The model accounts for the impact of supplied capital on lending rates and admits closed-form solutions under three interest-rate models: linear, kinked and adaptive. We then backtest the allocation strategy on Morpho lending markets on Ethereum, considering USDC first and WETH second.

The backtests emphasize the \textit{dilution effect} faced by lenders, which is the counterpart of the borrower-side \textit{size effect} studied in \citet{baude2026leveraged}. On the borrower side, larger leveraged positions increase utilization and raise borrowing costs, reducing profitability. On the lender side, larger supplied positions reduce utilization and dilute the lending rates, also reducing profitability. On both sides, the rate impact of positions is thus central to the optimization problem.

The kink also plays an important role. We show that an allocation that brings a market exactly to the kink is never optimal on the lender side, whereas it can be optimal on the borrower side \citep{baude2026leveraged}. This asymmetry may create tension between lenders and borrowers around the kink and hence exacerbate rate volatility.

A natural extension is to move beyond the single-agent setting. With many lenders, optimal allocations would jointly affect utilization and hence the lending rates earned by all lenders. With both lenders and borrowers, optimal lending and leveraged strategies would interact through utilization, affecting both lending and borrowing rates. This leads to a Nash equilibrium problem, where individually optimal strategies may be collectively suboptimal. This is an important direction for future research.

\bibliography{bibliographie.bib}

\section*{Appendix}

\appendix
\addtocontents{toc}{\protect\setcounter{tocdepth}{-5}}
\renewcommand*{\thesubsection}{\Alph{subsection}}

\subsection{Proof of Proposition \ref{pr:linear_rate_solution}}\label{appendix:proof_linear_rate_solution}

By plugging \eqref{eq:linear_rate_model} into the objective function in Problem \eqref{eq:baseline_problem}, we obtain:
\begin{equation}\label{eq:linear_problem}
\begin{aligned}
x^{\ast} = \underset{x}{\mathrm{argmax}} \quad & x_{0} r + \sum^{n}_{i=1} f_{i} (x_{i}) \\
\textrm{s.t.} \quad & \sum^{n}_{i=0} x_{i} = \xi \\
                    & x_{0} \geq 0 \\
                    & x^{\text{min}}_{i} \leq x_{i} \leq x^{\text{max}}_{i}, \quad i=1,\ldots,n
\end{aligned}
\end{equation}
where,
\begin{equation}\label{eq:f_linear}
f_{i} (x_{i}) = r_{base} \frac{\bar{B}_{i} x_{i}}{\bar{S}_{i} + x_{i}} + \frac{r_{slope1}}{u^{*}} \frac{\bar{B}^{2}_{i} x_{i}}{(\bar{S}_{i} + x_{i})^{2}}
\end{equation}

By introducing a Lagrange multiplier $\lambda$ for the equality $\sum^{n}_{i=0} x_{i} = \xi$, we obtain:
\begin{equation}\label{eq:linear_problem_Lagrange}
\begin{aligned}
x^{\ast}(\lambda) = \underset{x}{\mathrm{argmax}} \quad & x_{0} (r - \lambda) + \sum^{n}_{i=1} f_{i} (x_{i}) + \lambda \big ( \xi - \sum^{n}_{i=1} x_{i} \big ) \\
\textrm{s.t.} \quad & x_{0} \geq 0 \\
                    & x^{\text{min}}_{i} \leq x_{i} \leq x^{\text{max}}_{i}, \quad i=1,\ldots,n
\end{aligned}
\end{equation}

From the constraint $x_{0} \geq 0$, the KKT conditions imply directly:
\begin{equation}
\lambda \geq r, \quad x_{0} (\lambda - r) = 0
\end{equation}
Hence, if $x^{\ast}_{0} > 0$, then $\lambda^{\ast} = r$, else if $\lambda^{\ast} > r$, then $x^{\ast}_{0} = 0$. In any cases, $\lambda \geq r > 0$.

For a fixed $\lambda > 0$, the problem separates across markets. Hence, for each market $i$, we have to solve the one-dimensional problem:
\begin{equation}\label{eq:one_dimension_problem}
\underset{x^{\text{min}}_{i} \leq x_{i} \leq x^{\text{max}}_{i}}{\mathrm{max}} \quad g_{i} (x_{i}, \lambda)
\end{equation}
where,
\begin{equation}\label{eq:g_linear}
g_{i} (x_{i}, \lambda) = f_{i} (x_{i}) - \lambda x_{i}
\end{equation}
The purpose of the following analysis is to show that, this one-dimensional problem admits a unique maximizer, even though $f_{i}$ is not globally concave. The first derivative of $f_{i}$ is:
\begin{equation}\label{eq:f_linear_prime}
f'_{i} (x_{i}) = r_{base} \frac{\bar{B}_{i} \bar{S}_{i}}{(\bar{S}_{i} + x_{i})^{2}} + \frac{r_{slope1}}{u^{*}} \frac{\bar{B}^{2}_{i} (\bar{S}_{i} - x_{i})}{(\bar{S}_{i} + x_{i})^{3}}
\end{equation}
and the second derivative of $f_{i}$ is:
\begin{equation}\label{eq:f_linear_prime_prime}
f''_{i}(x_{i}) = 2 \frac{ \big ( \frac{r_{slope1}}{u^{*}} \bar{B}_{i}^{2} - r_{base} \bar{B}_{i} \bar{S}_{i} \big ) x_{i} - \bar{S}_{i} \big ( r_{base} \bar{B}_{i} \bar{S}_{i} + 2 \frac{r_{slope1}}{u^{*}} \bar{B}_{i}^{2} \big )}{(\bar{S}_{i} + x_{i})^{4}}
\end{equation}

As we assume $r_{base} < r_{slope1} \frac{\bar{u}_{i}}{u^{*}}$, the function $f_{i}$ is concave first and then convex for large $x_{i}$. However, the first derivative of $f_{i}$ vanishes at the unique point:
\begin{equation}\label{eq:peak_point}
x^{(\text{peak})}_{i} = \bar{S}_{i} \frac{r_{slope1} \frac{\bar{u}_{i}}{u^{*}} + r_{base}}{r_{slope1} \frac{\bar{u}_{i}}{u^{*}} - r_{base}}
\end{equation}

Since $\lambda > 0$, we have, for $x_{i} \geq x^{(\text{peak})}_{i}$:
\begin{equation}
\frac{\partial g_{i}}{\partial x_{i}} (x_{i}, \lambda) = f'_{i} (x_{i}) - \lambda < 0
\end{equation}
Therefore, $g_{i} (\cdot, \lambda)$ is strictly decreasing on $[x^{(\text{peak})}_{i}, +\infty)$ and no point strictly larger than $x^{(\text{peak})}_{i}$ can be an interior maximizer. Therefore, any interior maximizer of $g_{i} (\cdot, \lambda)$, if it exists, must belong to $[0, x^{(\text{peak})}_{i}]$. It remains to study $g_{i} (\cdot, \lambda)$ on this interval.

The function $f_{i}$ is strictly concave on $[0, x^{(\text{infl})}_{i}]$, where,
\begin{equation}\label{eq:inflection_point}
x^{(\text{infl})}_{i} = \bar{S}_{i} \frac{2 r_{slope1} \frac{\bar{u}_{i}}{u^{*}} + r_{base}}{r_{slope1} \frac{\bar{u}_{i}}{u^{*}} - r_{base}} > x^{(\text{peak})}_{i}
\end{equation}
Thus $f_{i}$ is strictly concave on $[0, x^{(\text{peak})}_{i}]$ and consequently $g_{i} (\cdot, \lambda)$ is also strictly concave on $[0, x^{(\text{peak})}_{i}]$. The FOC consequently identifies at most one interior candidate, and this candidate is the unique interior maximizer. The FOC reads:
\begin{equation}\label{eq:FOC}
r_{base} \frac{\bar{B}_{i} \bar{S}_{i}}{(\bar{S}_{i} + x_{i})^{2}} + \frac{r_{slope1}}{u^{*}} \frac{\bar{B}^{2}_{i} (\bar{S}_{i} - x_{i})}{(\bar{S}_{i} + x_{i})^{3}} = \lambda
\end{equation}

By considering the following change of variable: $y_{i} = x_{i} + \bar{S}_{i}$ we obtain:
\begin{equation}
y^{3}_{i} + p_{i} y_{i} + q_{i} = 0
\end{equation}
where,
\begin{equation}
p_{i} = \frac{\bar{B}_{i} \bar{S}_{i}}{\lambda} \Big ( r_{slope1} \frac{\bar{u}_{i}}{u^{*}} - r_{base} \Big ), \quad q_{i} = - 2 r_{slope1} \frac{\bar{B}_{i}^{2} \bar{S}_{i}}{\lambda u^{*}}
\end{equation}
Then, the discriminant $\Delta$ is given by:
\begin{equation}
\Delta(p_{i}, q_{i}) = - \big ( 4 p^{3}_{i} + 27 q^{2}_{i} \big ) < 0
\end{equation}
because $p_{i} > 0$ and $q_{i} < 0$ and we obtain the unique solution of the interior maximizer $\kappa(p_{i}, q_{i})$ using Cardano's formula.

Now we incorporate the bounds: $x^{\text{min}}_{i} \leq x_{i} \leq x^{\text{max}}_{i}$ and the KKT conditions gives us the optimal solution:
\begin{equation}
x^{\ast}_{i}(\lambda) = \left\{
    \begin{array}{ll}
        \displaystyle x^{\text{min}}_{i} & \mbox{if } f'(x^{\text{min}}_{i}) < \lambda \\
        \displaystyle \kappa(p_{i}, q_{i}) - \bar{S}_{i} & \mbox{if } f'(x^{\text{max}}_{i}) \leq \lambda < f'(x^{\text{min}}_{i}) \\
        \displaystyle x^{\text{max}}_{i} & \mbox{if } f'(x^{\text{max}}_{i}) > \lambda
    \end{array}
\right.
\end{equation}
with, $f'(x^{\text{max}}_{i}) < f'(x^{\text{min}}_{i})$ provided: $x^{\text{max}}_{i} < x^{(\text{infl})}_{i}$. Equivalenty, we can reads:
\begin{equation}
x^{\ast}_{i}(\lambda) = \min \big ( \max \big ( \kappa(p_{i}, q_{i}) - \bar{S}_{i}, x^{\text{min}}_{i} \big ), x^{\text{max}}_{i} \big )
\end{equation}

\subsection{Proof of Proposition \ref{pr:kinked_rate_solution} and Corollary \ref{co:kink_non_optimal}}\label{appendix:proof_kinked_rate_solution}

For fixed $\lambda > 0$, the kinked rate splits the feasible interval into two regions separated by $x^{(\text{kink})}_{i} = \frac{\bar{B}_{i}}{u^{*}} - \bar{S}_{i}$. On each region, the borrow rate is linear in utilization; hence the corresponding restriction of $g(\cdot, \lambda)$ is of the same form as in the linear-rate case and admits a unique branch-wise maximizer. Since $g(\cdot, \lambda)$ is continuous at the kink and the feasible interval is closed and bounded, a global maximizer exists. Moreover, by the standard finite-piece maximization argument, any global maximizer must be either one of the branch-wise maximizers or the kink point itself. Thus the solution is obtained by computing the two branch-wise candidates, adding the kink point when feasible, and selecting the candidate with the largest value of $g(\cdot, \lambda)$. If this largest value is attained by a single candidate, the maximizer is unique.

The kinked rate function is piecewise linear and has a discontinuity in its derivative at $u^{*}$. Its subdifferential at the kink is the interval between the left and right derivatives:
\begin{equation}
\partial b_{i}(u^{*}) = \Big [ \frac{r_{slope1}}{u^{*}}, \frac{r_{slope2}}{1-u^{*}} \Big ]
\end{equation}
Then, we have:
\begin{equation}
\lambda^{1}_{i} = u^{*} \Big (\frac{u^{*}}{\bar{u}_{i}} (r_{base} + r_{slope1}) + r_{slope1} \big ( \frac{u^{*}}{\bar{u}_{i}} - 1 \big ) \Big )
\end{equation}
\begin{equation}
\lambda^{2}_{i} = u^{*} \Big ( \frac{u^{*}}{\bar{u}_{i}} (r_{base} + r_{slope1}) + \frac{r_{slope2} u^{*}}{1-u^{*}} \big ( \frac{u^{*}}{\bar{u}_{i}} - 1 \big ) \Big )
\end{equation}
Therefore, when $\lambda \in [\lambda^{1}_{i}, \lambda^{2}_{i}]$, the optimal solution is $x^{(\text{kink})}_{i}$. However, as $\lambda^{1}_{i} > \lambda^{2}_{i}$ since $\frac{r_{slope2}}{1-u^{*}} > \frac{r_{slope1}}{u^{*}}$ then the interval is empty. As a conclusion, $x^{(\text{kink})}_{i}$ is never optimal.

Additionally, whenever $\lambda_{i}^{1} < r$, the pre-kink regime can be discarded. Indeed, the presence of the external fixed-rate opportunity implies $\lambda^{*} \geq r$, whereas the pre-kink regime can only be reached for $\lambda \le \lambda_{i}^{1}$.

\subsection{Optimal allocation on WETH markets}\label{appendix:weth}

\begin{table}
\centering
\begin{tabular}{ccc}
\toprule
market & ID & creation date \\
\midrule
WETH/wstETH & \texttt{6becf9b4-3c85-40bf-9938-196812e034a3} & March 14, 2024 \\
WETH/weETH & \texttt{1e21b2df-853b-4ed7-b26f-72f0feb1708f} & September 23, 2024 \\
WETH/WBTC & \texttt{9797d4d3-5c72-48b7-a87d-d720ef1a02b4} & March 29, 2024 \\
WETH/cbBTC & \texttt{22f29890-ffb5-4bb3-bc9a-dfc1b2f103b4} & September 4, 2024 \\
\bottomrule
\end{tabular}
\caption{IDs and creation dates of the WETH markets on Morpho on the Ethereum blockchain.}
\label{tab:weth_markets}
\end{table}
We consider the four largest WETH lending markets in terms of supplied liquidity as of January 1, 2026; their characteristics are summarized in Table \ref{tab:weth_markets}.

\begin{figure}
    \centering
    \begin{subfigure}{\linewidth} 
        \centering
        \includegraphics[scale = 0.6]{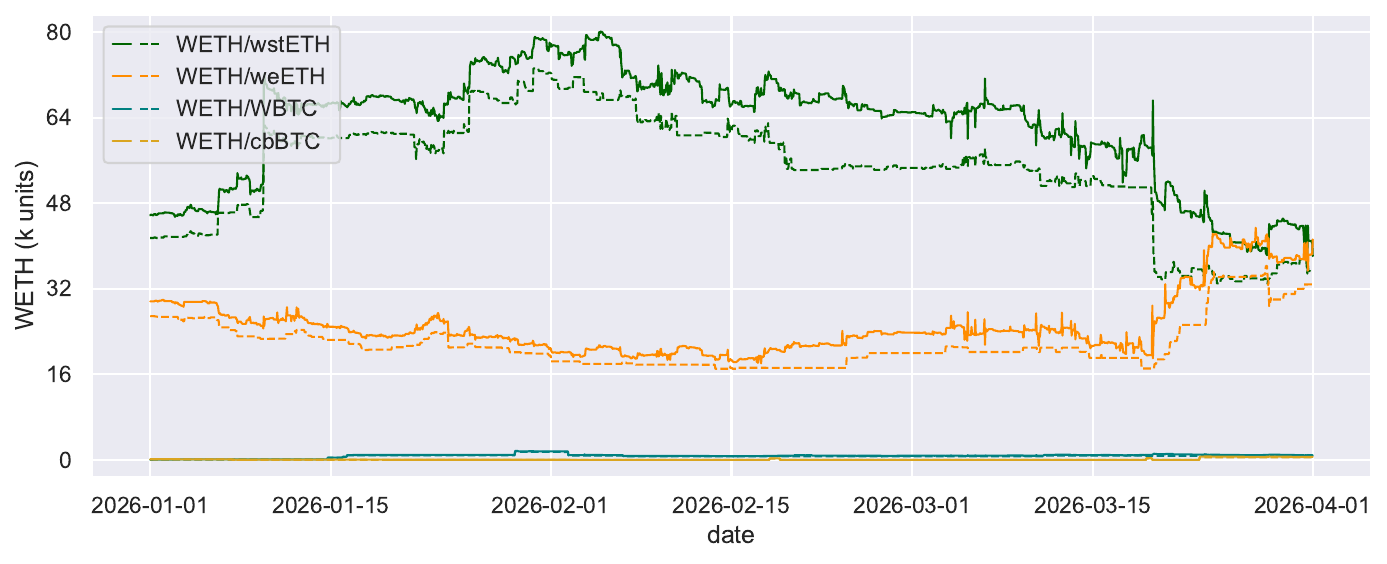}
        \caption{reserves}
    \end{subfigure}
    \vfill
    \begin{subfigure}{\linewidth}
        \centering
        \includegraphics[scale = 0.6]{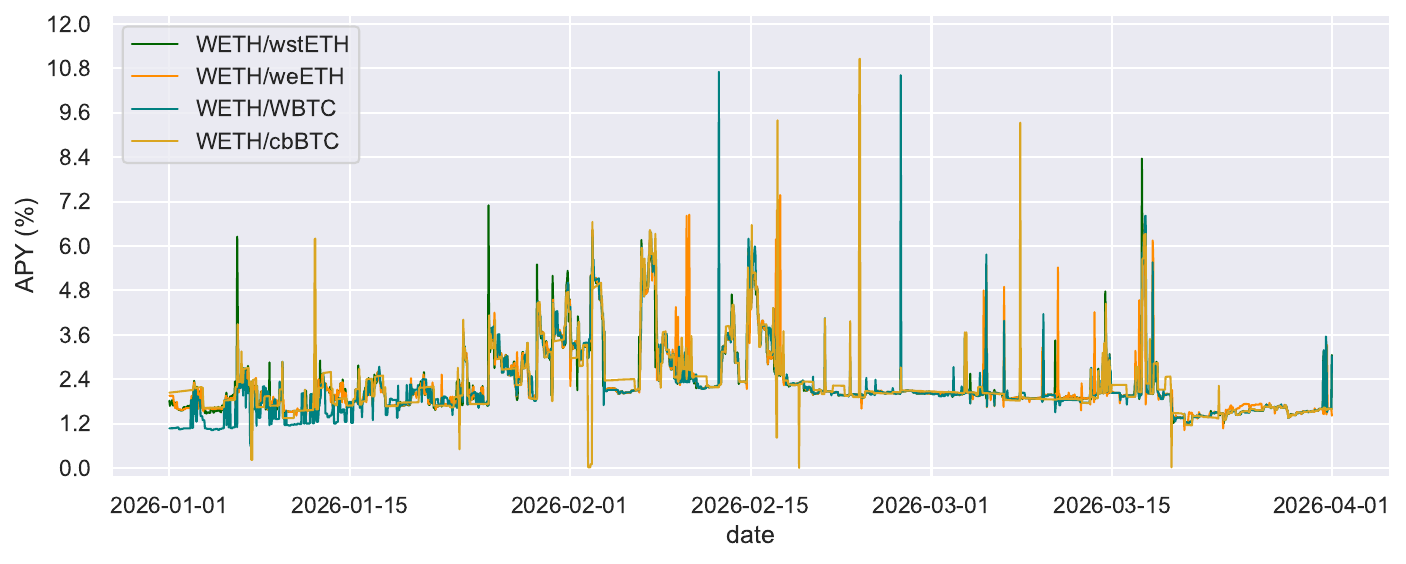}
        \caption{interest rate}
    \end{subfigure}
    \caption{Evolution of WETH reserves (solid line: supplied funds; dashed line: borrowed fund) and the interest rate of the WETH markets on Morpho on the Ethereum blockchain from January 1, 2026 to April 1, 2026.}
    \label{fig:evolution_weth_market_rates}    
\end{figure}
Figure \ref{fig:evolution_weth_market_rates} shows the evolution of supplied and borrowed WETH, together with the corresponding supply rates, over the backtesting period. As in the core-inspired USDC market set, supply rates are relatively close across markets. At the beginning of the backtest, the WETH/wstETH market is substantially more liquid than the other selected markets, but its liquidity later becomes comparable to that of the WETH/weETH market.

\begin{figure}
    \centering
    \begin{subfigure}{\linewidth} 
        \centering
        \includegraphics[scale = 0.6]{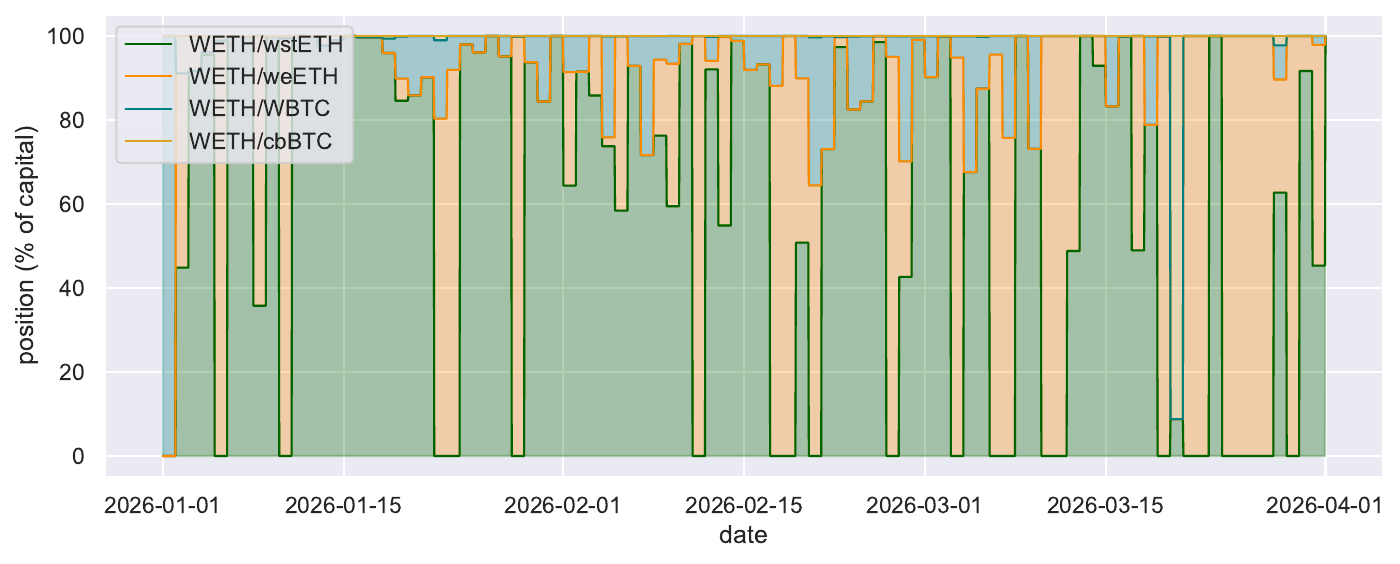}
        \caption{low cap}
    \end{subfigure}
    \vfill
    \begin{subfigure}{\linewidth}
        \centering
        \includegraphics[scale = 0.6]{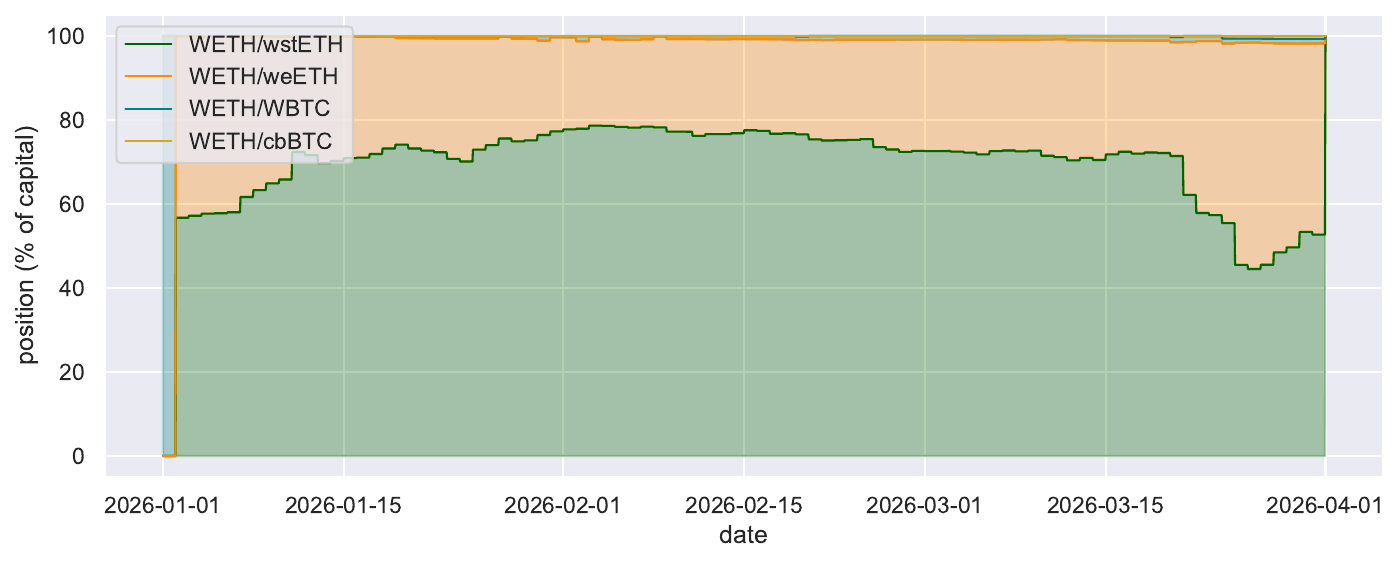}
        \caption{high cap}
    \end{subfigure}
    \caption{Evolution of the positions of the WETH-based strategy on Morpho on the Ethereum blockchain from January 1, 2025 to April 1, 2025.}
    \label{fig:backtesting_ethereum_weth}    
\end{figure}
\begin{figure}
    \centering
    \includegraphics[scale = 0.6]{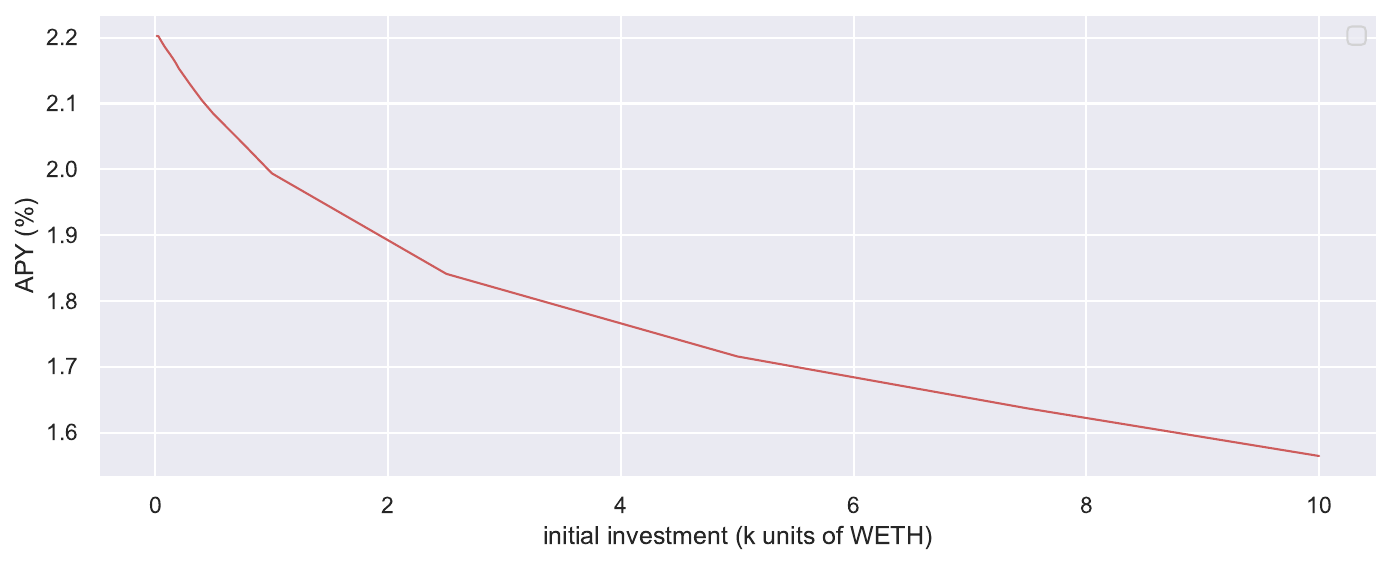}
    \caption{APY of the WETH-based strategy with respect to initial investment on Morpho on the Ethereum blockchain from January 1, 2026 to April 1, 2026.}
    \label{fig:backtesting_ethereum_weth_apys}
\end{figure}

Figure \ref{fig:backtesting_ethereum_weth} reports the resulting allocations for the low- and high-capital cases, while Figure \ref{fig:backtesting_ethereum_weth_apys} shows the APY as a function of the initial investment. Since liquidity is mainly concentrated in the WETH/wstETH and WETH/weETH markets, large allocations are mostly absorbed by these two markets. Consequently, the \textit{dilution effect} is primarily driven by the aggregate liquidity of these dominant markets.

\end{document}